%% file: main.tex
\documentclass[conference,10pt]{IEEEtran}
\IEEEoverridecommandlockouts
\usepackage[switch]{lineno}
\usepackage{cite}
\usepackage{fancyhdr}
\usepackage[table,xcdraw]{xcolor}
\usepackage[normalem]{ulem}

\usepackage[font={small, color={black}}]{caption}
\usepackage{subcaption}

\usepackage{listings}
\usepackage{courier}
\usepackage{graphicx}
\usepackage{arydshln}
\usepackage{url}
\usepackage{hyperref}
\hypersetup{
 colorlinks=true,
 linkcolor=blue,
 filecolor=blue,
 citecolor = blue,      
 urlcolor=blue,
}

\usepackage{enumitem}
\setlist[itemize]{noitemsep, topsep=0.5pt, parsep=0pt, partopsep=0pt}
\setlist[enumerate]{noitemsep, topsep=1pt, parsep=0pt, partopsep=0pt}

\usepackage[numbers]{natbib}
\usepackage{multicol}
\definecolor{codegreen}{rgb}{0,0.6,0}
\definecolor{codegray}{rgb}{0.5,0.5,0.5}
\definecolor{codepurple}{rgb}{0.58,0,0.82}
\definecolor{backcolour}{rgb}{0.95,0.95,0.92}

\lstdefinestyle{mystyle}{
    backgroundcolor=\color{backcolour},   
    commentstyle=\color{magenta},
    keywordstyle=\color{magenta},
    numberstyle=\tiny\color{codegray},
    stringstyle=\color{codepurple},
    basicstyle=\ttfamily\footnotesize,
    breakatwhitespace=false,         
    breaklines=true,                 
    captionpos=b,                    
    keepspaces=true,                 
    numbers=left,                    
    numbersep=5pt,                  
    showspaces=false,                
    showstringspaces=false,
    showtabs=false,                  
    tabsize=2
}
\usepackage{xcolor}

\newcommand{\tcolB}{\textcolor{blue}}

\definecolor{navyblue}{rgb}{0.0, 0.0, 0.5}

\usepackage{amssymb,amsfonts}

\usepackage{amsmath}
\usepackage{graphicx}
\usepackage{booktabs}
\usepackage{textcomp}
\usepackage{amsmath,amssymb,amsfonts}
\usepackage{graphicx}
\usepackage{textcomp}
\usepackage{xcolor}
\usepackage{etoolbox}
\usepackage{algorithm}
\usepackage{algpseudocode}
\usepackage{lipsum}
\usepackage{perpage} 
\MakePerPage{footnote} 

\usepackage{csquotes}

\newcolumntype{C}[1]{>{\centering\arraybackslash}p{#1}}
\newcolumntype{P}[1]{>{\raggedright\arraybackslash}p{#1}}
\def\BibTeX{{\rm B\kern-.05em{\sc i\kern-.025em b}\kern-.08em
    T\kern-.1667em\lower.7ex\hbox{E}\kern-.125emX}}

\begin{document}

\title{Transfer learning for conflict and duplicate
detection in software requirement pairs
}
\author{\IEEEauthorblockN{1\textsuperscript{st} Garima Malik}
\IEEEauthorblockA{
\textit{Toronto Metropolitan University}\\
Toronto, Canada \\
garima.malik@torontomu.ca}
\and
\IEEEauthorblockN{2\textsuperscript{nd} Savas Yildirim}
\IEEEauthorblockA{
\textit{Toronto Metropolitan University}\\
Toronto, Canada \\
savas.yildirim@torontomu.ca}
\and
\IEEEauthorblockN{3\textsuperscript{nd} Mucahit Cevik}
\IEEEauthorblockA{
\textit{Toronto Metropolitan University}\\
Toronto, Canada \\
mcevik@torontomu.ca}
}

\maketitle

\begin{abstract}
Conflicting and duplicate software requirements can lead to design errors, increased development cost, and project delays if they are not identified early. However, automatically detecting such inconsistencies is difficult because software requirements are written in natural language and often contain complex semantics. In this work, we formulate conflict and duplicate detection as a requirement pair classification task and propose \textit{SR-BERT}, a transformer-based framework that combines bi-encoder architectures with domain adaptation. To evaluate the proposed approach, we construct two proprietary datasets, namely CDN and CN, with the help of software development experts and transform four publicly available requirement datasets into requirement pair classification datasets. We then investigate sequential multi-stage fine-tuning and cross-domain transfer learning strategies across these datasets. Experimental results show that sequential fine-tuning consistently improves performance over baseline models, with SR-BERT achieving the best results on larger datasets. Cross-domain experiments further demonstrate that the proposed framework generalizes well across both industrial and open-source requirements. Overall, our findings show that domain-adapted transformer models provide an effective solution for automating software requirement conflict and duplicate detection.
\end{abstract}

\begin{IEEEkeywords}
Software requirements, conflict and duplicate detection, transformer models, transfer learning, cross-domain
\end{IEEEkeywords}

\section{Introduction}
Requirement Engineering (RE) is a fundamental phase of software development that focuses on identifying, documenting, and managing software requirements. These requirements are usually captured in a Software Requirements Specification (SRS) document, which describes the expected system functionalities and constraints. To ensure successful software development, the requirements should be accurate, consistent, and complete. Poor-quality requirements can lead to design errors, increased development costs, project delays, and reduced user satisfaction \citep{wiegers2013software,lamsweerde2009requirements,lamsweerde2009requirements}. However, SRS documents often contain inconsistencies such as conflicting requirements, where one requirement contradicts another, and duplicate requirements, where the same requirement is expressed using different wording. If these issues are not detected early, they can negatively affect software quality and system reliability \citep{guo2021automatically,malik2025supervised,robertson2012mastering}.

Automatically detecting conflicting and duplicate requirements remains challenging for several reasons. First, conflicting requirement pairs are relatively rare, resulting in severe class imbalance during model training \citep{malik2025supervised}. Second, software requirements often contain domain-specific terminology and depend on project context, making it difficult to determine whether two requirements truly conflict \citep{guo2021automatically}. Third, the same requirement can be written in many different ways, with variations in wording and sentence structure, making duplicate detection challenging \citep{pohl2016requirements}. Finally, software requirements evolve throughout the development process, and requirements that are initially consistent may become conflicting as the system changes \citep{urbieta2012detecting}.

Recent advances in Natural Language Processing (NLP), especially pretrained transformer models such as BERT \citep{devlin2018bert} and RoBERTa \citep{liu2019roberta}, have significantly improved performance on a wide range of sentence classification tasks through transfer learning. In particular, the Natural Language Inference (NLI) task, which determines whether one sentence entails, contradicts, or is neutral with respect to another, closely resembles the problem of analyzing requirement pairs. Motivated by this similarity, we formulate conflict and duplicate detection as a requirement pair classification task and further investigate a binary conflict vs neutral formulation for conflict detection.

The main contributions of this work is as follows:
\begin{itemize}
    \item We formulate conflict and duplicate identification as a requirement pair classification task and curate two expert-annotated datasets, \textsc{CDN} (Conflict, Duplicate, and Neutral) and \textsc{CN}, constructed following industry guidelines with requirements authored by software engineering professionals.
    
    \item We demonstrate the effectiveness of sequential transfer learning by fine-tuning BERT checkpoints first on large-scale NLI datasets and subsequently on requirement pair classification data, establishing a strong baseline for the task.

    \item We propose SR-BERT, a domain-adapted bi-encoder architecture that leverages pretrained Sentence-BERT and further adapts it to the software requirements domain through unsupervised domain adaptation, enabling more accurate semantic encoding of requirement pairs.

    \item We systematically evaluate cross-domain transfer learning to assess the generalizability of BERT-based checkpoints across diverse requirement datasets, providing insights into their capability in real-world deployment scenarios.
\end{itemize}

\section{Literature Review}\label{sec:background}
\subsection{NLP and Transfer Learning in Requirements Engineering}
Natural Language Processing (NLP) has become central to automating requirements engineering (RE) tasks, including requirement classification, quality assessment, and information extraction \citep{nazir2017applications,zhao2021natural}. Recent work has shifted from lexical and syntactic features toward deep embedding techniques, demonstrating strong performance across semantic-level RE tasks \citep{sonbola2022use}. Transfer learning, particularly through pretrained transformer models such as BERT \citep{devlin2018bert} and DeBERTa \citep{he2021deberta}, has further advanced the field by reducing dependence on large labeled datasets while improving generalization \citep{ruder2019}.
In RE specifically, transfer learning has been successfully applied to requirement classification \citep{hey2020norbert,kici2021bert}, anaphoric ambiguity detection \citep{ezzini2022taphsir}, and domain concept extraction \citep{ajagbe2022retraining}. Studies have shown that domain-adapted language models (e.g., NoRBERT, BERT4RE) outperform generic pretrained models on software requirement texts \citep{hey2020norbert,ajagbe2022retraining}. However, industrial validation remains limited due to scarce public RE datasets \citep{zhao2021natural}.

\subsection{Sentence Pair Classification}
Sentence pair classification underpins many NLP tasks relevant to RE, including semantic textual similarity (STS) \citep{reimers2019sentence}, paraphrase identification \citep{xu2015semeval}, and NLI \citep{maccartney2008}. Transformer-based models, fine-tuned on large NLI datasets such as Multi-Genre NLI (MNLI) \citep{williams2017broad}, have proven particularly effective for determining semantic relationships between text pairs. The label structure and scale of MNLI (433k instances) make it especially suitable for transfer learning to domain-specific pair classification tasks, including requirement comparison.

\subsection{Conflict and Duplicate Identification}
Automated detection of conflicting requirements remains a challenging problem in Requirement Engineering (RE). Existing approaches can be broadly grouped into four categories. The first category uses rule-based techniques that combine Natural Language Processing (NLP) pipelines with manually designed heuristics to identify conflicts \citep{ghosh2014automatically,guo2021automatically,aldekhail2017intelligent}. The second category applies fuzzy logic to detect and resolve requirement conflicts \citep{polpinij2008automatic,viana2017identifying}. The third category employs deep learning models, particularly BiLSTM-based architectures, to identify conflicts in non-functional requirements \citep{abeba2021identification}. More recently, transformer-based methods have been proposed, combining semantic similarity with entity overlap analysis to improve conflict detection \citep{malik2025supervised}.

Compared with conflict detection, duplicate requirement detection has received much less attention. Although duplicate detection has been widely studied for bug reports using CNN and RNN-based models \citep{he2020duplicate,wang2020duplicate}, similar research on software requirements is limited. Existing studies mainly focus on empirical analyses, and there is still a lack of well-established benchmarks and robust learning-based approaches for identifying duplicate requirements \citep{falessi2011empirical,rao2023research}.

\section{Methodology}
\subsection{Datasets}
\begin{figure*}[!t]
     \centering
     \subfloat[WorldVista\label{fig:w_v_cosine}]{
     \includegraphics[width=0.282\textwidth]{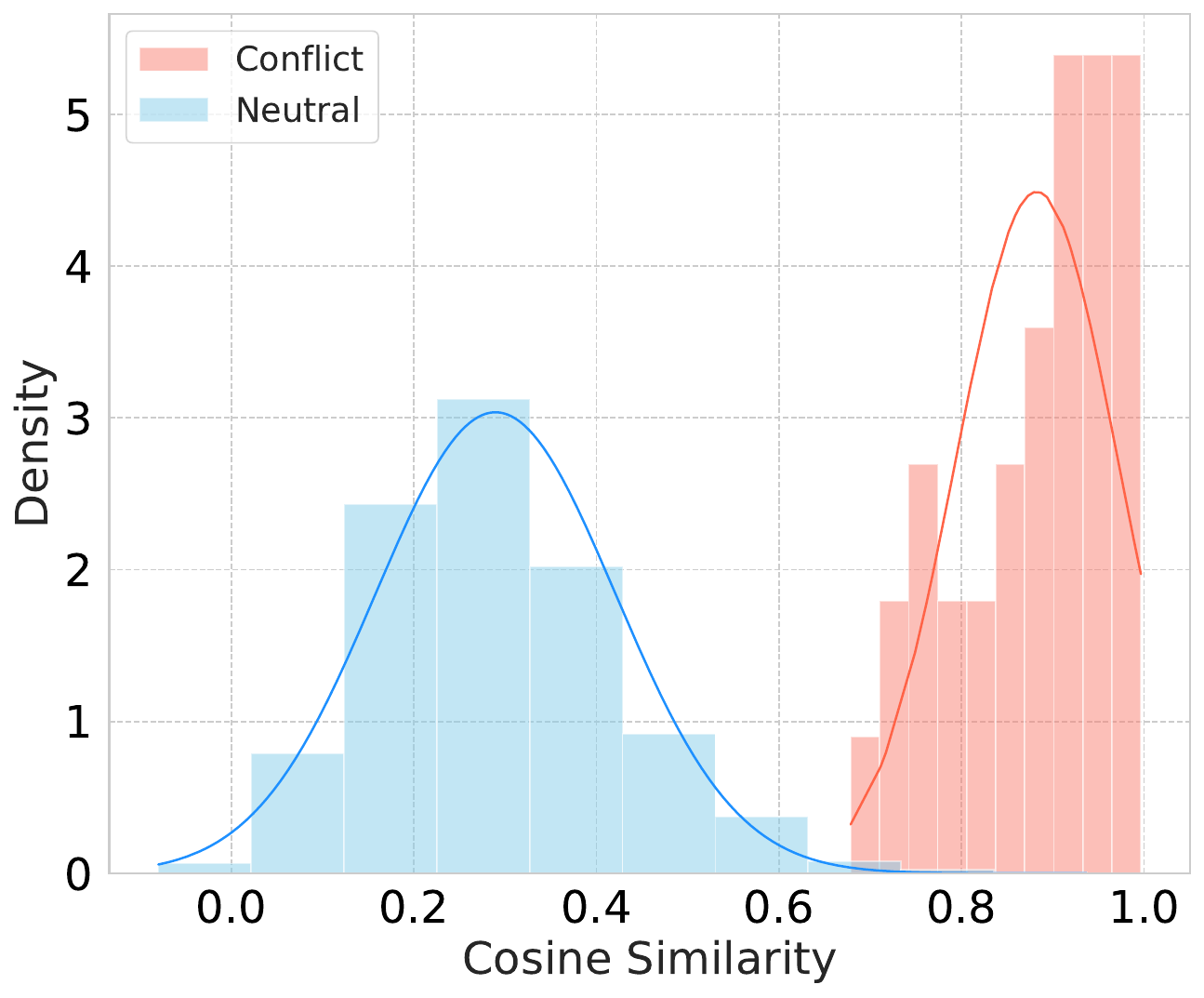}}\hfill
     \subfloat[UAV\label{fig:uav_cosine}]{\includegraphics[width=0.282\textwidth]{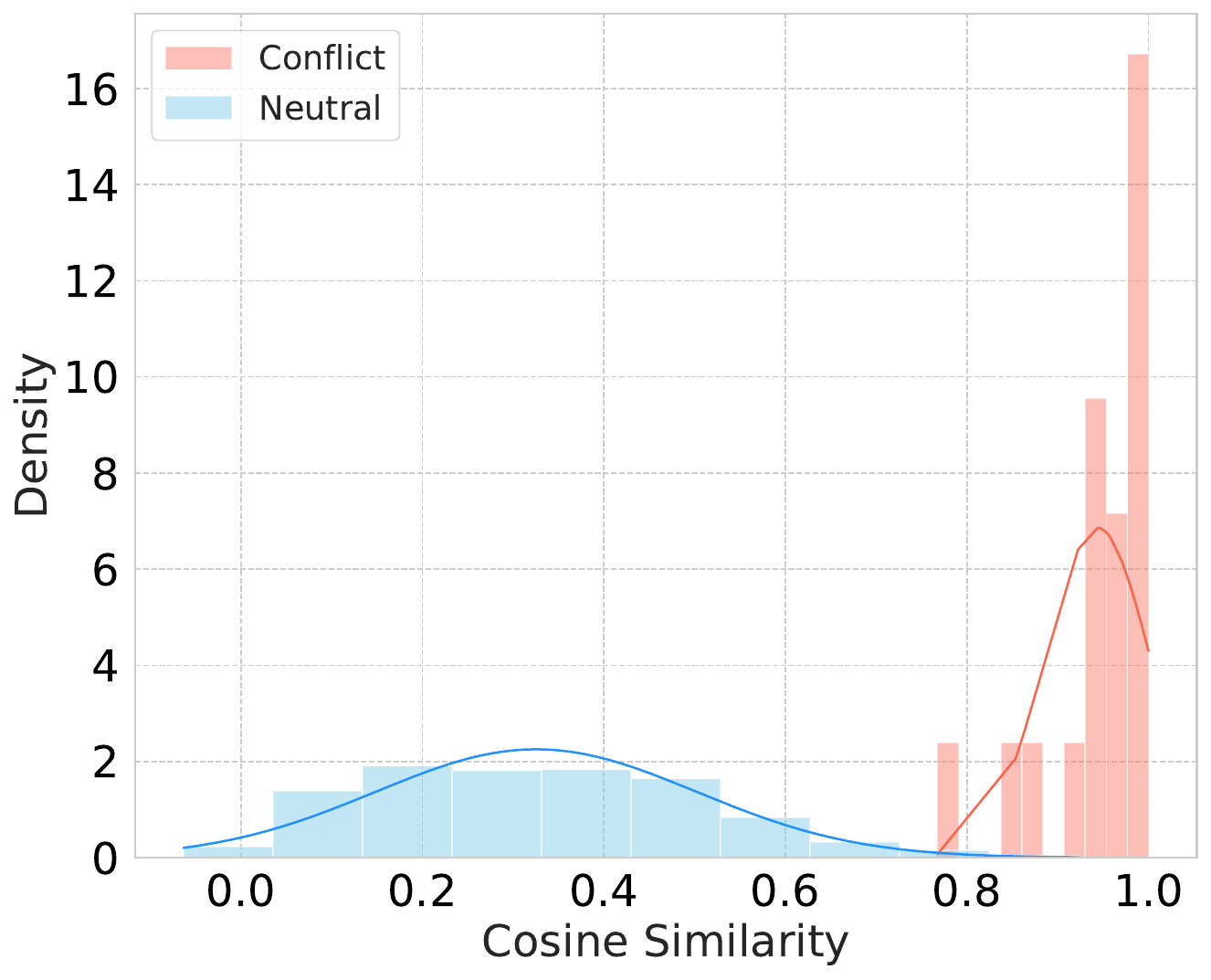}}\hfill
     \subfloat[PURE\label{fig:pure_cosine}]{\includegraphics[width=0.302\textwidth]{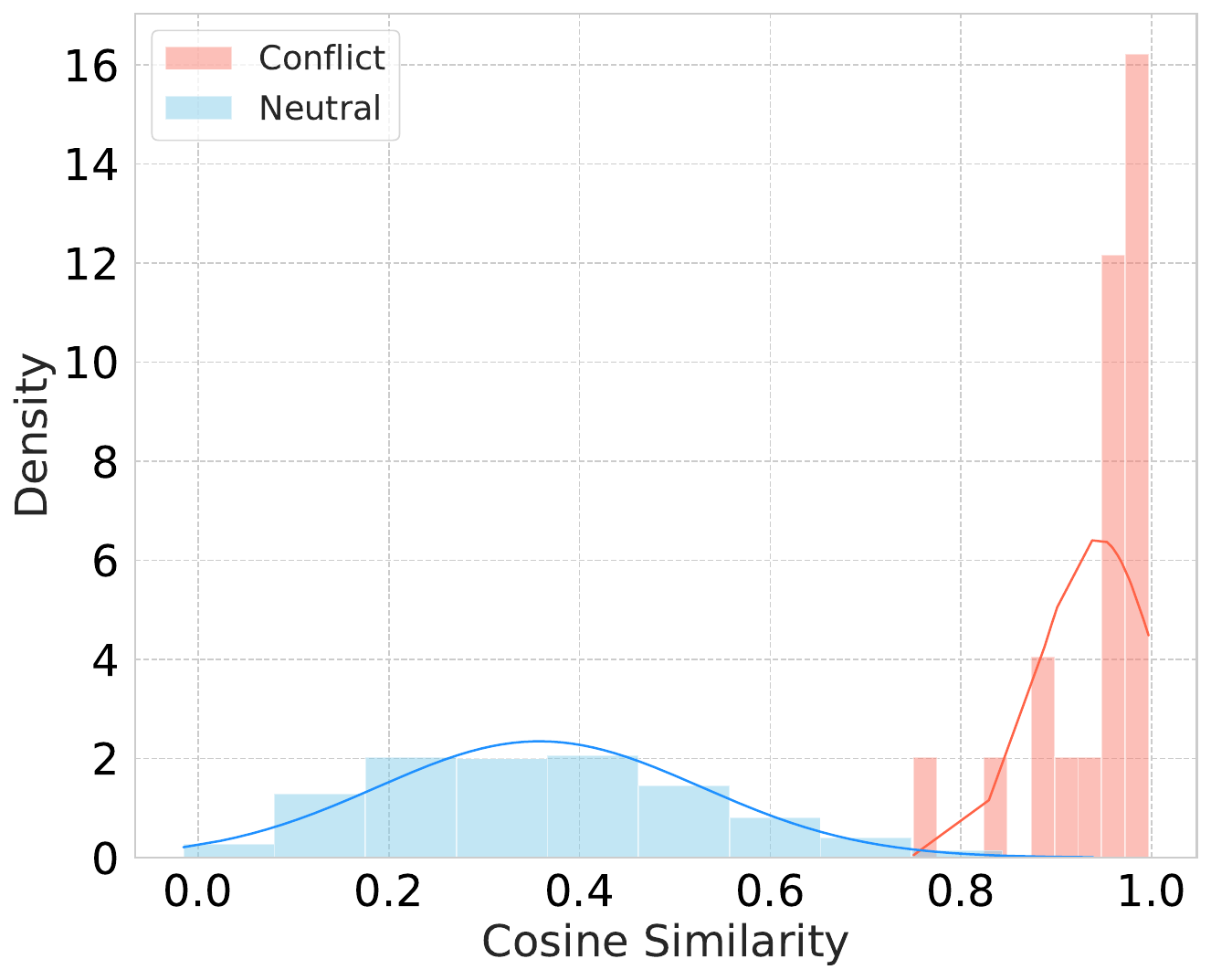}} \\
     \subfloat[OPENCOSS\label{fig:open_cosine}]{\includegraphics[width=0.282\textwidth]{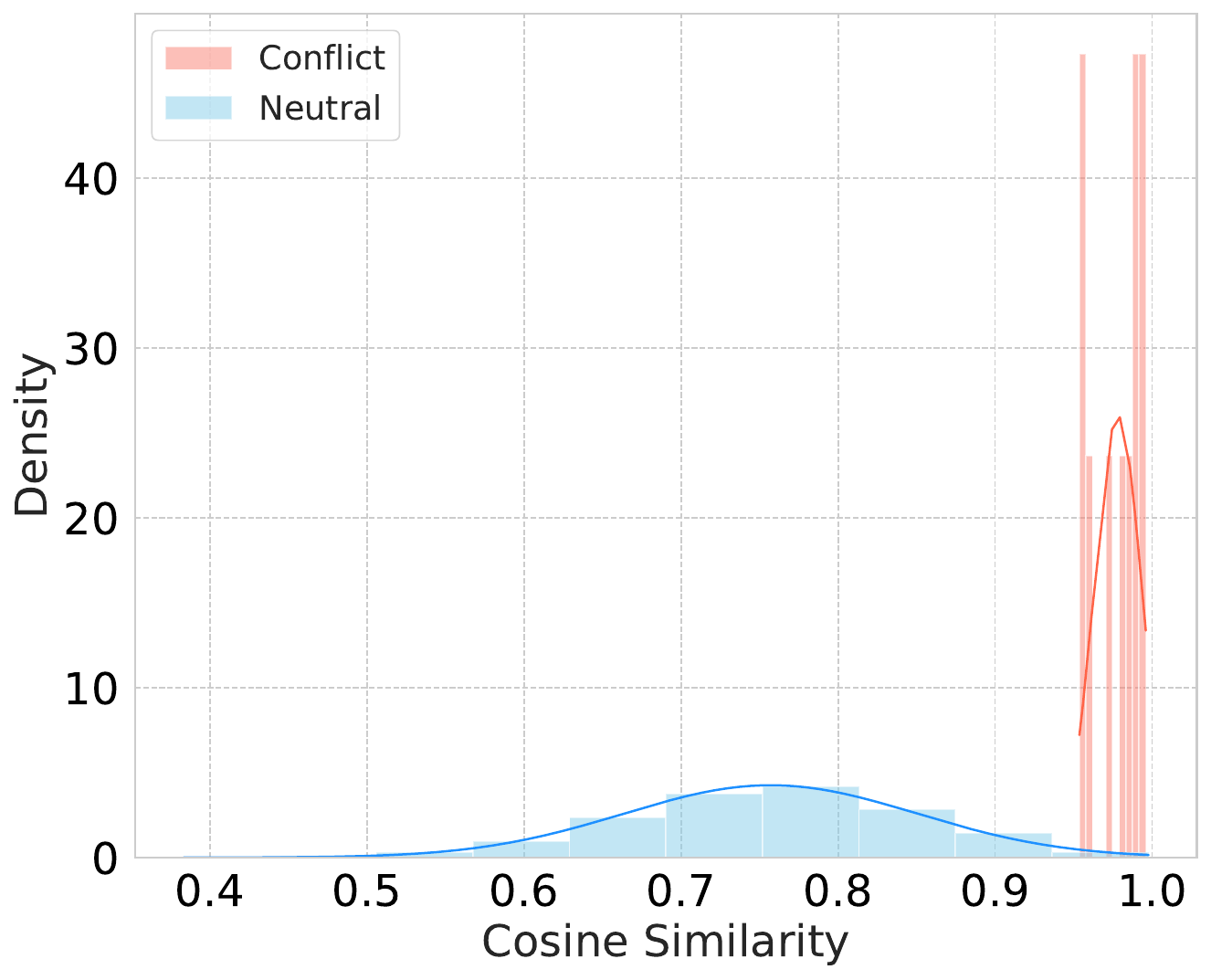}}\hfill
     \subfloat[CDN\label{fig:cdn_cosine}]{\includegraphics[width=0.282\textwidth]{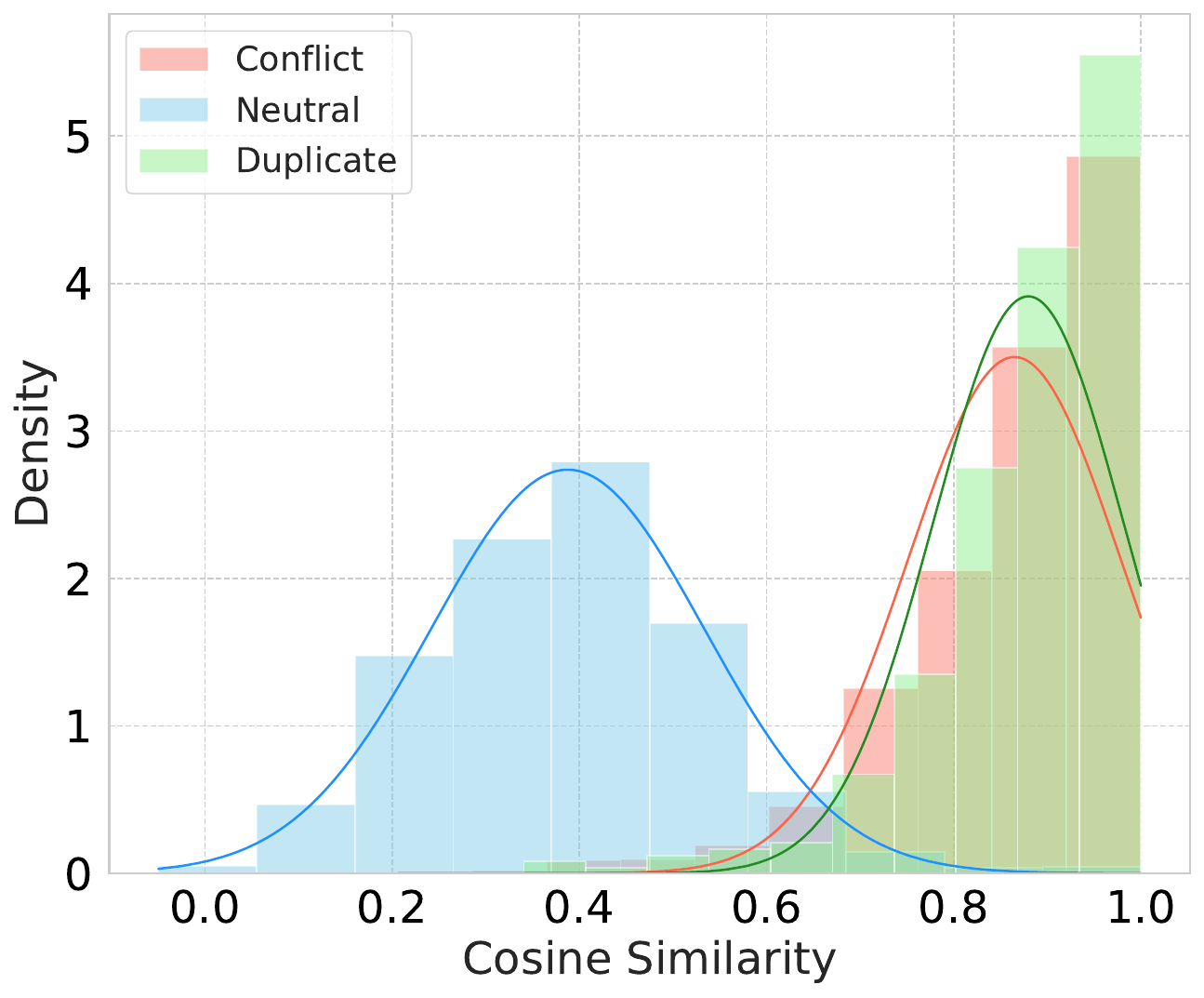}}\hfill
     \subfloat[CN\label{fig:cn_cosine}]{\includegraphics[width=0.282\textwidth]{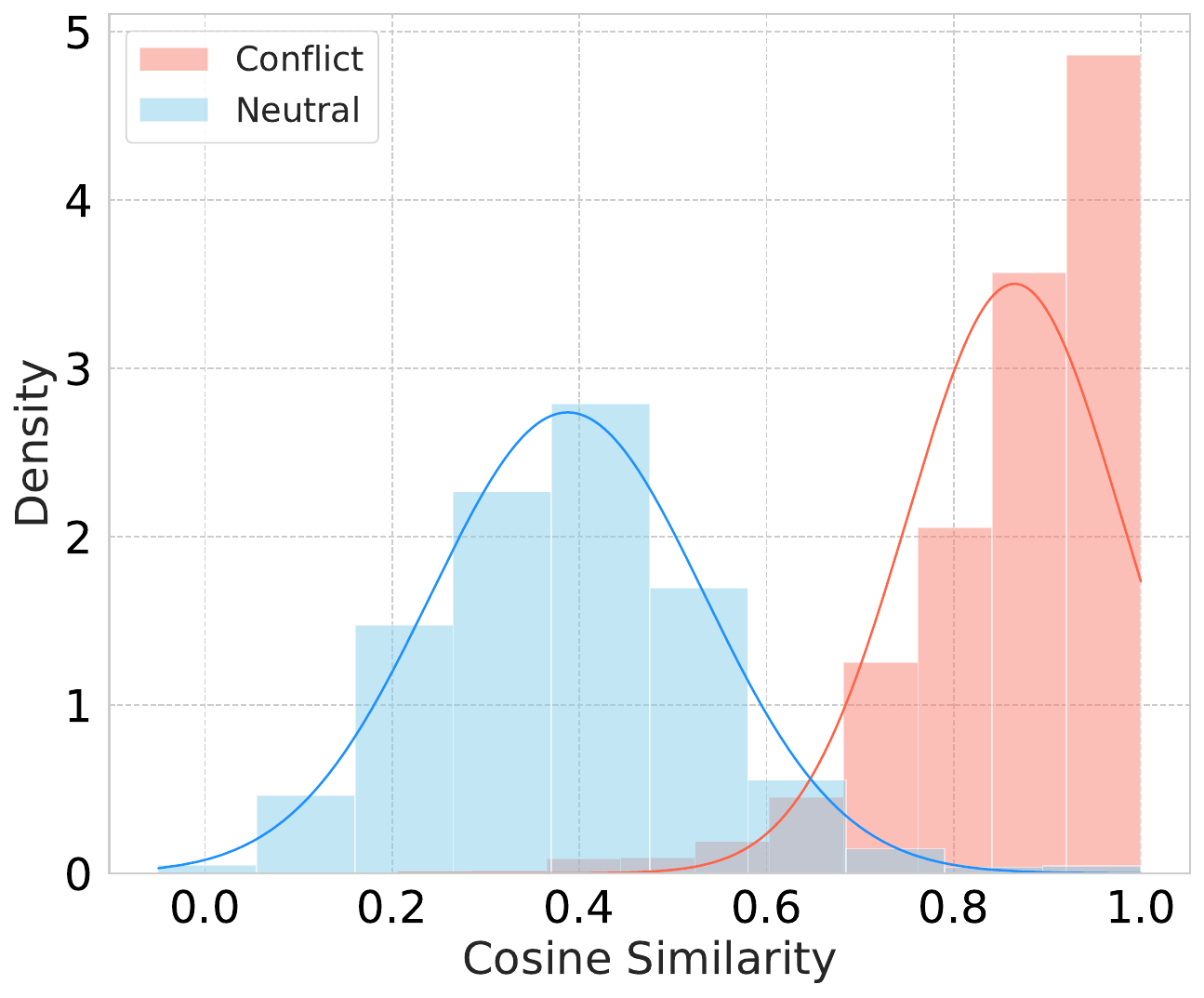}}
     \caption{Cosine similarity distribution of requirement pairs for the respective class labels, computed using SBERT embeddings. 
     }
     \label{fig:type_merged}
 \end{figure*}

We evaluate our approach on six datasets spanning software engineering, healthcare, transportation, and hardware: two proprietary datasets (CDN and CN) and four open-source SRS datasets (UAV, WorldVista, PURE, and OPENCOSS). When requirements are written as long paragraphs, we segment them into individual sentences to obtain a consistent requirement-pair representation. We first describe the proprietary datasets and then summarize the open-source datasets used in the cross-domain experiments.
\begin{itemize}
    \item \textit{CDN}: Derived from IBM-DOORS requirements and user stories \citep{doors}, comprising three labels: Conflict, Duplicate, and Neutral. A team of 10 IBM software developers created conflict requirements for 48 original requirements, while nine developers composed duplicates. Domain experts verified all pairs. To mitigate class imbalance, neutral pairs were limited to half the sum of conflict and duplicate pairs. 
    Table~\ref{tab:data_sample} illustrates the pair structure. For example, requirement pair (1,2) is labeled as a conflict because the modifier ``only'' and the alternative command sources (``pilot controller'' versus ``remote viewing app'') imply incompatible access rules. Duplicate pairs instead use synonyms or closely related phrases, whereas neutral pairs are drawn from distinct domains or SRS documents with different actors and actions. Table~\ref{tab:class_dist} reports the class distribution, and Figure~\ref{fig:cdn_cosine} shows the cosine-similarity distributions. The overlap between conflict and duplicate pairs in this embedding space illustrates why the three-class task is nontrivial.
\setlength{\tabcolsep}{7.5pt}
\renewcommand{\arraystretch}{1.15}
\begin{table*}[!t]
\centering
    \caption{Example data instances from the CDN dataset, showing requirement pairs and their respective labels, with emphasis on words and phrases that indicate the assigned label.}
    \label{tab:data_sample}
    \resizebox{\linewidth}{!}{
\begin{tabular}{lp{0.35\textwidth} p{0.35\textwidth} p{0.08\textwidth}}
\toprule
\textbf{Req. pair Id} &\textbf{Requirement Text 1} & \textbf{Requirement Text 2} & \textbf{Label} \\ 
\midrule
(1,2) &The UAV shall \textbf{\tcolB{only}} accept commands from an authenticated \textbf{\tcolB{Pilot controller}}. & The UAV shall accept commands from an authenticated \textbf{\tcolB{remote viewing app}}. & Conflict\\
(3,4) &All UAV data transmission shall be secure against \textbf{\tcolB{unauthorized access}}. &
All UAV data transmission shall be secure against \textbf{\tcolB{all access}}. &
Conflict \\
\midrule
(5,6) & These are the \textbf{\tcolB{client}} requirements for the Aviary System of Systems. & These are the \textbf{\tcolB{customer}} requirements for the Aviary System. & Duplicate\\
(7,8) & The UAV shall send the Pilot real-time information about \textbf{\tcolB{malfunctions}} that \textbf{\tcolB{impact}} the mission. & 
The UAV shall send the Pilot real-time data about \textbf{\tcolB{errors}} that \textbf{\tcolB{affect}} the mission. & Duplicate \\
\midrule
(9,10) & The \textbf{\tcolB{Pilot}} shall be able to \textbf{\tcolB{fly}} the UAV to any accessible-by-air location within 20 miles of its origin & The \textbf{\tcolB{UAV}} shall \textbf{\tcolB{charge}} to 75\% in less than 3 hours. & Neutral\\
(11,12) & The \textbf{\tcolB{UAV}} shall periodically \textbf{\tcolB{send}} the Pilot the power and estimated flight time remaining. &
A \textbf{\tcolB{single adult}} shall be able to \textbf{\tcolB{lift}} and carry the UAV. & Neutral \\
\bottomrule
\end{tabular}
}
\end{table*}

\item \textit{CN}: We also use a binary variant, CN, derived from CDN by removing duplicate pairs to enable cross-domain transfer experiments. Figure~\ref{fig:cn_cosine} shows that conflict pairs exhibit higher cosine similarity than neutral pairs, as conflicting requirements often share similar verbs and actors (Table~\ref{tab:data_sample}). This structure makes CN a useful source dataset for testing whether conflict-relevant semantic patterns transfer to other requirement corpora.
\end{itemize}

We obtained the UAV, WorldVista, PURE, and OPENCOSS datasets from \citet{malik2025supervised}, this paper utilized these datasets to evaluate the supervised semantic similarity-based conflict detection approach (S3CDA). These datasets initially contained limited conflict instances; to enrich them, \citet{malik2025supervised} supplemented each with synthetic conflict requirements crafted following INCOSE guidelines to align with industry standards. Table~\ref{tab:class_dist} summarizes the class distribution, average token length, and vocabulary size for each dataset.
\begin{itemize}
\item \textit{UAV}: Requirements for a UAV control system from the University of Notre Dame, structured using the EARS template \citep{mavin2009easy}. Figure~\ref{fig:uav_cosine} shows clear separation in cosine similarity between neutral and conflict labels.

\item \textit{WorldVista}: Health management system requirements\footnote[2]{https://worldvista.org/Documentation} with basic healthcare terminology. This dataset contains 35 conflict and 10,843 neutral pairs. Figure~\ref{fig:w_v_cosine} shows similar distinct cosine similarity distributions across labels.

\item \textit{PURE}: Derived from two SRS documents (THEMAS and Mashbot) in the public PURE corpus \citep{pure}. Long paragraphs were segmented into individual sentences. The dataset contains 20 conflict and 2,191 neutral pairs. Figure~\ref{fig:pure_cosine} shows cosine similarity patterns comparable to the UAV dataset.

\item \textit{OPENCOSS}: Railway and automotive safety requirements\footnote{http://www.opencoss-project.eu}. With only 10 conflict pairs against 6,776 neutral pairs, Figure~\ref{fig:open_cosine} exhibits a heavily skewed conflict label distribution.
\end{itemize}

\begin{table}[!t]
\centering
 \caption{Dataset statistics for requirement pair datasets. 
 }
    \label{tab:class_dist}
    \resizebox{\linewidth}{!}{
\begin{tabular}{c|rrrrrr}
\toprule
{\bf Dataset} & {\bf \# Conflict}  & {\bf \# Duplicate} & {\bf \# Neutral} & {\bf Avg. \# tokens in pairs} & {\bf Vocabulary size}\\
\midrule
CDN  & 5,553 & 1,673 & 3,400 & (18.19, 18.43) & 1022  \\
\cmidrule(lr){2-6}
WorldVista &35 &- & 10,843 &(23.09, 21.80) & 884 \\
UAV &18 &- &6,652& (26.44, 26.30) & 298 \\
PURE &20 &- &2,191 & (26.44, 26.30) & 298 \\
OPENCOSS &10 &- & 6,776 &(29.85, 30.02) & 250 \\
CN & 5,553 &- &3,400 & (18.23, 18.51) & 1022 \\
\bottomrule
\end{tabular}
}
\end{table}
\subsection{Sequential Transfer Learning}
\begin{figure}[!t]
    \centering
    \includegraphics[width=0.5\textwidth]{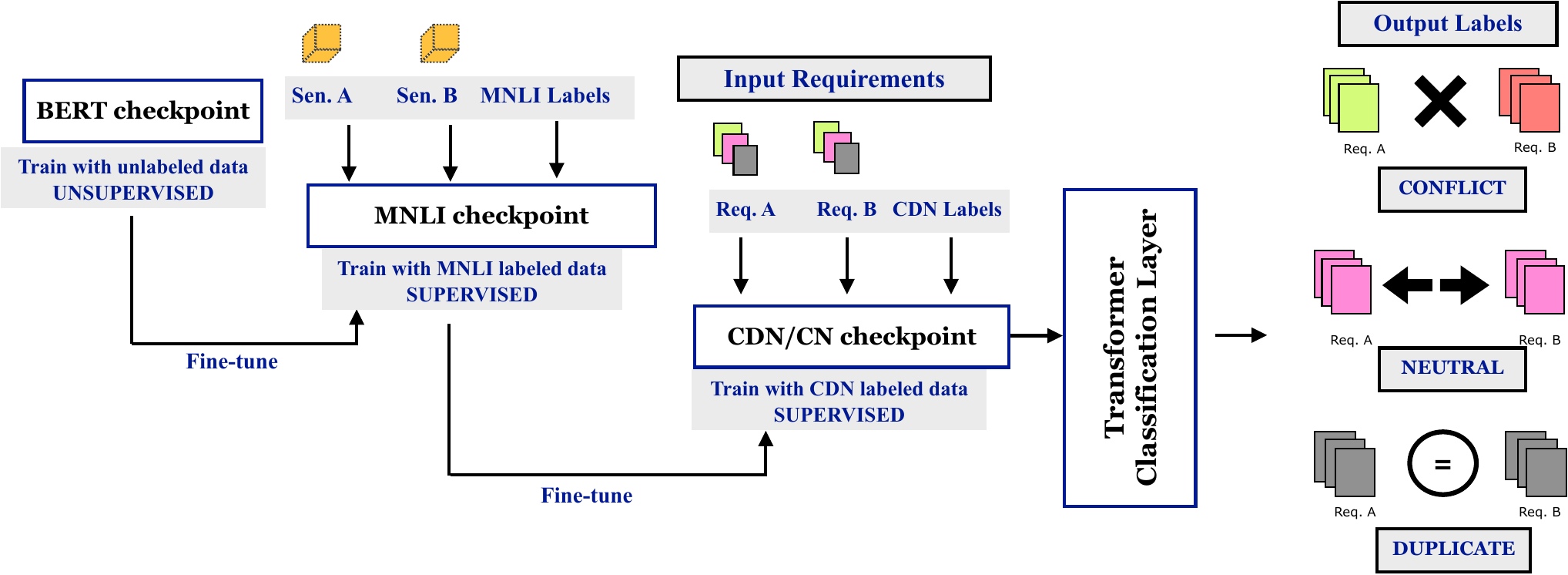}
    \caption{A visual description of the sequential transfer learning approach for requirement pair classification}
    \label{fig:method_3}
\end{figure}

Sequential transfer learning leverages knowledge from a source domain to improve target task performance by progressively adapting language model checkpoints \citep{ruder2019}. While transfer learning is often described as pretraining followed by task-specific fine-tuning, recent work also considers multi-stage adaptation through intermediate labeled or unlabeled datasets, provided negative transfer is avoided \citep{ruder2019}. 
Our experiments use both direct task fine-tuning and multi-stage supervised fine-tuning. Figure~\ref{fig:method_3} illustrates the sequential pipeline: a vanilla BERT checkpoint is first fine-tuned on the MNLI dataset to learn general sentence-pair inference patterns and is then fine-tuned on requirement-pair datasets such as CDN to adapt those patterns to conflict and duplicate detection.
To isolate the value of the intermediate NLI stage, we compare these sequential checkpoints with vanilla BERT variants that are fine-tuned directly on the requirement-pair datasets.

\subsection{SR-BERT}

Sentence-BERT (SBERT) is an extension of BERT designed to produce semantically meaningful sentence embeddings while enabling efficient sentence-level comparisons \citep{reimers2019sentence}. Unlike the original BERT architecture, which jointly processes sentence pairs during inference, SBERT adopts a bi-encoder architecture that independently encodes each sentence into a fixed-length vector. This design significantly reduces the computational cost of pairwise comparison and makes SBERT well suited for tasks such as semantic similarity, information retrieval, and sentence pair classification.

SBERT is trained using siamese and triplet network architectures, in which multiple copies of the same encoder share parameters during training. The shared-weight design encourages the model to learn a representation space where semantically related sentences are placed closer together while unrelated sentences are pushed farther apart.

\begin{figure}[!t]
    \centering
    \includegraphics[width=0.5\textwidth]{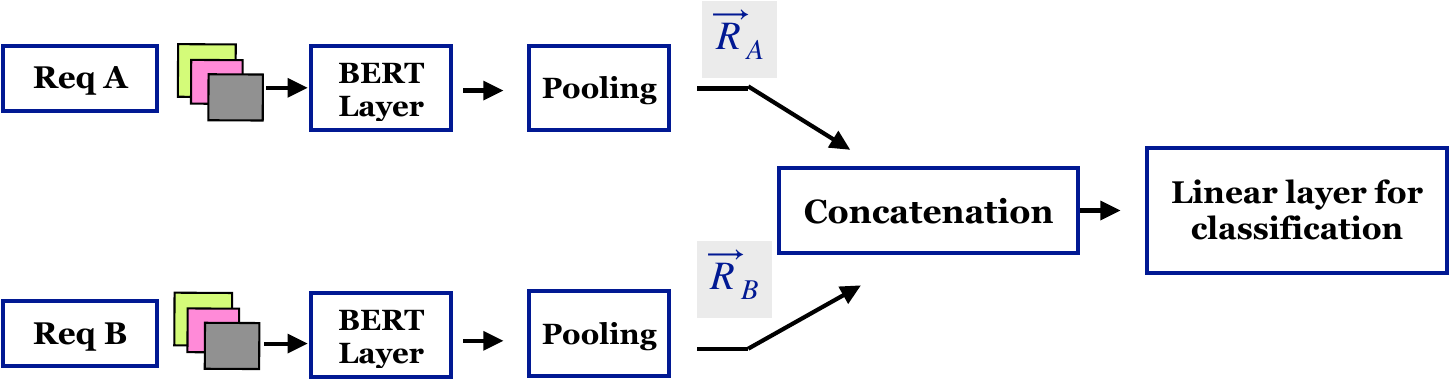}
    \caption{SBERT-based model for software requirement pair classification. Individual requirements are encoded using SBERT to obtain sentence embeddings, which are subsequently concatenated and passed through a linear classifier.}
    \label{fig:sbert_embed}
\end{figure}

Figure~\ref{fig:sbert_embed} illustrates the use of SBERT for software requirement pair classification. Each requirement is independently encoded to obtain its sentence embedding, after which the embeddings are combined and passed to a linear classifier for prediction.

SBERT can be trained using either siamese or triplet learning objectives.
\begin{itemize}
    \item \textit{Siamese networks}: Two identical encoder networks independently process a pair of input sentences while sharing the same parameters. The objective is to minimize the embedding distance between semantically similar sentence pairs and maximize the distance between dissimilar pairs.
    \item \textit{Triplet networks}: Three identical encoder networks process an anchor sentence, a semantically similar (positive) sentence, and a semantically dissimilar (negative) sentence. The model learns embeddings by simultaneously bringing the anchor closer to the positive example while pushing it away from the negative example.
\end{itemize}
The triplet learning objective is defined as:
\begin{align}
L = \max \left(0,\,
D(\mathbf{R}_A,\mathbf{R}_P) -
D(\mathbf{R}_A,\mathbf{R}_N) + \epsilon
\right),
\end{align}
where $\mathbf{R}_A$, $\mathbf{R}_P$, and $\mathbf{R}_N$ denote the embeddings of the anchor, positive, and negative requirements, respectively, $D(\cdot,\cdot)$ represents a distance metric (typically cosine or Euclidean distance), and $\epsilon$ is the margin that enforces a minimum separation between positive and negative pairs. For example,
\begin{itemize}
\item $R_A$: \textit{The system shall allow users to reset their passwords.}
\item $R_P$: \textit{Users should be able to reset their passwords in the system.}
\item $R_N$: \textit{The software must provide real-time data synchronization.}
\end{itemize}
Here, $R_A$ and $R_P$ express the same functional requirement using different wording and therefore should be mapped to nearby locations in the embedding space. In contrast, $R_N$ represents an unrelated functionality and should be placed farther away from the anchor representation.

A key advantage of SBERT is that sentence embeddings can be generated independently and reused across multiple downstream tasks without repeatedly encoding sentence pairs. Consequently, SBERT serves as an effective feature extractor for transfer learning, where the generated embeddings can be employed for semantic similarity, clustering, retrieval, or classification tasks. Compared with cross-encoder models, bi-encoder architectures such as SBERT offer lower pairwise inference cost while preserving sentence-level semantic representations.

\begin{figure}[!t]
    \centering
    \includegraphics[width=0.5\textwidth]{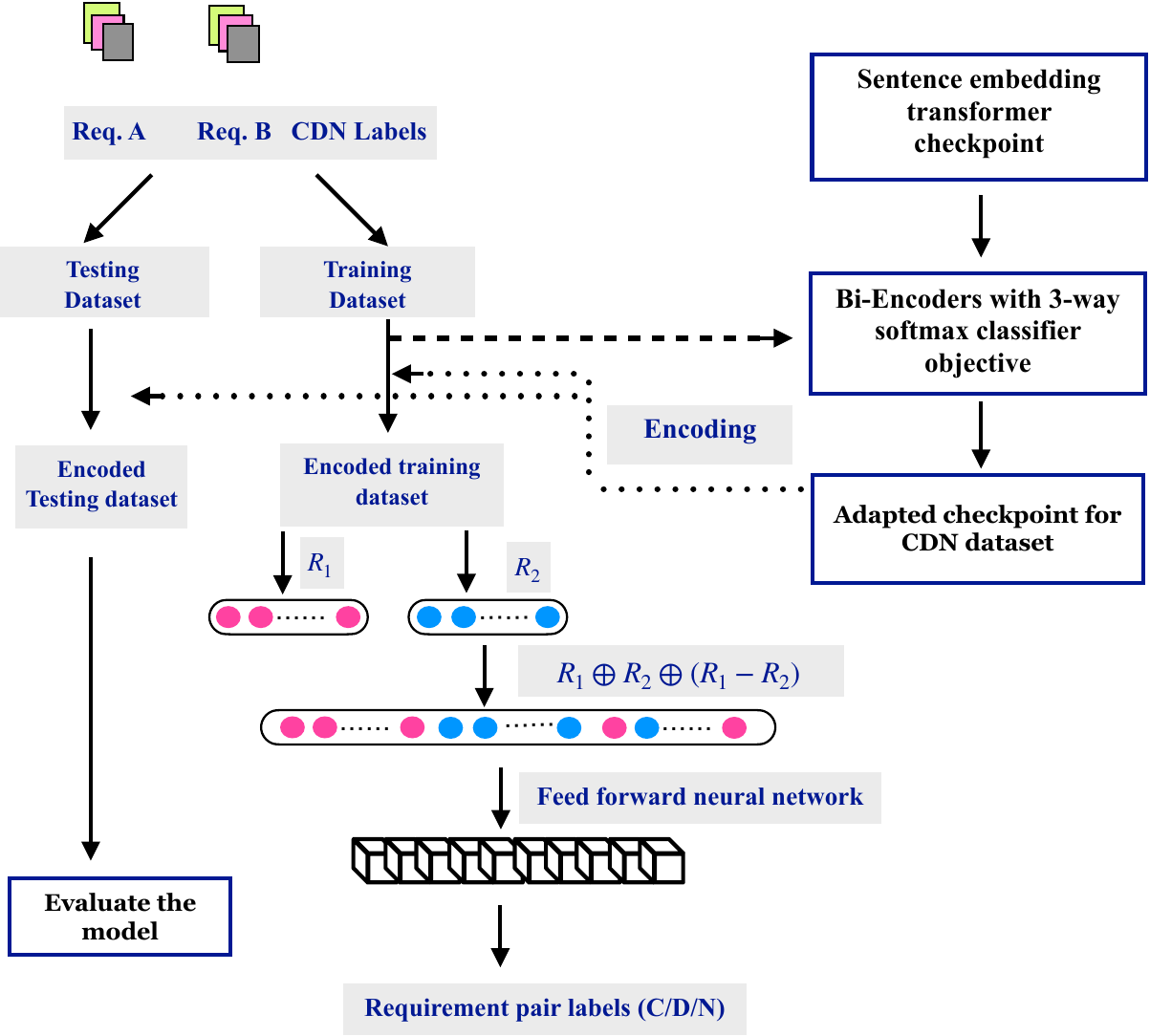}
    \caption{Architecture of the proposed SR-BERT model.}
    \label{fig:biencoder}
\end{figure}

Building upon these advantages, we propose \textit{Software Requirement BERT (SR-BERT)}, a domain-adapted SBERT framework for software requirement pair classification. As illustrated in Figure~\ref{fig:biencoder}, SR-BERT starts from a pretrained SBERT checkpoint and further fine-tunes it on labeled software requirement pairs using a supervised three-class classification objective. This domain adaptation enables the encoder to learn representations that better capture the semantic characteristics of software requirements.

Given a pair of requirements, the fine-tuned encoder generates two sentence embeddings, denoted by $R_1$ and $R_2$. Rather than relying solely on their concatenation, we construct the final pair representation as
\begin{align}
R = R_1 \oplus R_2 \oplus (R_1 - R_2),
\end{align}
where $\oplus$ denotes vector concatenation. The embeddings $R_1$ and $R_2$ preserve the semantic representation of each individual requirement, while the difference vector $(R_1-R_2)$ explicitly captures their semantic relationship. Incorporating both individual and relational information enables the classifier to distinguish requirement pairs with subtle semantic differences more effectively than using either representation alone.

The resulting feature vector is then passed through a feed-forward neural network with a softmax output layer, which is trained using the ground-truth labels to perform requirement pair classification.

\subsection{Baseline classification methods}
To evaluate the effectiveness of SR-BERT, we compare it against three representative baseline approaches that differ only in the sentence representation used while employing a comparable linear classifier.

\begin{itemize}
    \item \textit{TF-IDF + SVM}: Each requirement is represented using TF-IDF features. The feature vectors corresponding to a requirement pair are concatenated and classified using a linear Support Vector Machine (SVM).
    \item \textit{Word Embeddings (GloVe + FastText) + SVM}: Each requirement is represented by averaging pretrained GloVe and FastText word embeddings. The resulting sentence representations are concatenated and provided as input to a linear SVM classifier.
    \item \textit{SBERT + Linear Layer}: Requirements are independently encoded using a pretrained SBERT model. The resulting sentence embeddings are concatenated and fed directly to a linear classification layer, as illustrated in Figure~\ref{fig:sbert_embed}. Unlike SR-BERT, this baseline does not perform domain-specific adaptation of the SBERT encoder.

\end{itemize}
\subsection{Cross-domain transfer learning}
We use cross-domain transfer learning to test whether a transformer trained on one requirement-pair dataset can support classification in another dataset with the same label structure~\citep{ruder2019,cao2021deep, aderghal2018classification}. This experiment is motivated by shared requirement structure across the datasets and by the cosine-similarity patterns observed for conflict pairs. Formally, we define a source-domain dataset as $D_s = \left \{ (r_{s_1}^i,r_{s_2}^i),y_{s_{1,2}}^i \right \}_{i=1}^{N_s}$ and a target domain dataset as $D_t = \left \{ (r_{t_1}^i,r_{t_2}^i),y_{t_{1,2}}^i \right \}_{i=1}^{N_t}$, with $(r_{s_1},r_{s_2})$ representing a requirement pair in the source dataset and $y_{s_{1,2}}$ representing the label defining the relationship between requirement pairs. Similarly, $(r_{t_1},r_{t_2})$ represents a requirement pair in the target dataset, and $y_{t_{1,2}}$ represents the corresponding label.

\begin{figure}[!t]
    \centering
    \includegraphics[width=0.5\textwidth]{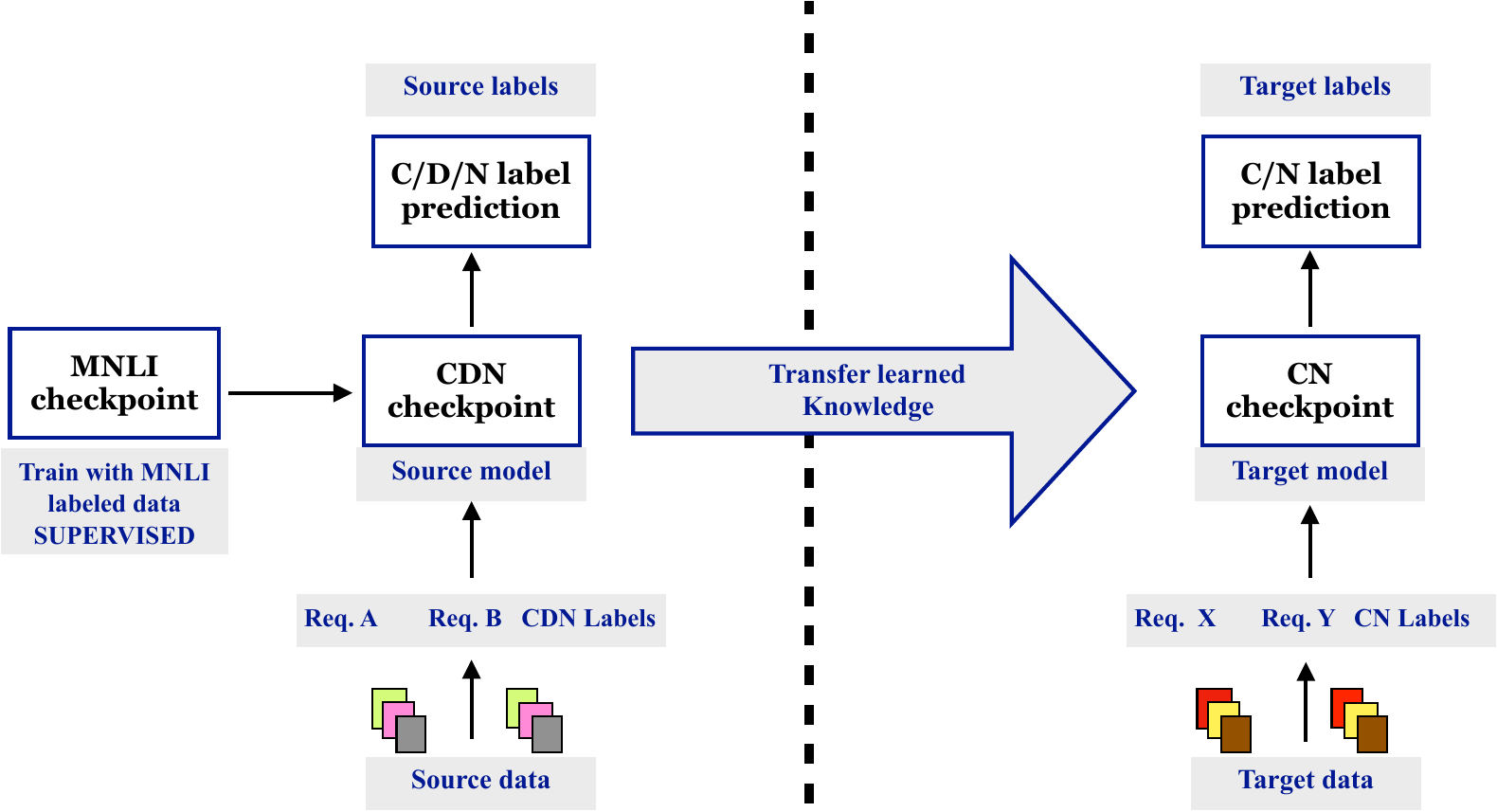}
    \caption{Cross-domain transfer learning framework for requirement pair classification task.}
    \label{fig:cross_domain}
\end{figure}

Figure~\ref{fig:cross_domain} illustrates the domain adaptation configuration for our experiments. In our requirement datasets, CDN and CN have relatively balanced class distributions, while the other requirement-pair datasets contain few conflict instances. We first train a transformer model on CN and evaluate it on the other requirement-pair datasets. We then train models on individual datasets and on combinations of UAV, WorldVista, OPENCOSS, and PURE, evaluating each model on the held-out dataset. This design allows us to assess whether shared structure in the input format and label definition is sufficient to support transfer across requirement sources.

\subsection{Experimental setup}
\label{sec:experiments}

We evaluate the proposed methods using several pretrained transformer models and their corresponding variants that have been further fine-tuned on the Multi-Genre Natural Language Inference (MNLI) dataset. Specifically, we consider three general-purpose pretrained language models, namely \textit{distilbert-base}\footnote[3]{https://huggingface.co/distilbert-base-uncased}, \textit{bert-base}\footnote{https://huggingface.co/bert-base-uncased}, and \textit{deberta-base}\footnote[5]{https://huggingface.co/microsoft/deberta-base}. In addition, we evaluate three MNLI-adapted checkpoints, namely \textit{roberta-large-mnli}\footnote[6]{https://huggingface.co/roberta-large-mnli}, \textit{bert-base-uncased-MNLI}\footnote[7]{https://huggingface.co/textattack/bert-base-uncased-MNLI}, and \textit{deberta-base-mnli}\footnote[8]{https://huggingface.co/microsoft/deberta-base-mnli}. These models are further fine-tuned on the software requirement pair datasets before evaluation.

Table~\ref{tab:parameters} summarizes the architectural characteristics of the transformer checkpoints together with the training hyperparameters adopted in our experiments.

\renewcommand{\arraystretch}{1.5}
\begin{table}[!t]
\centering
\caption{Transformer model checkpoint characteristics and hyperparameter details}
\label{tab:parameters}
\resizebox{\linewidth}{!}{
\begin{tabular}{lrrrrrr}
\toprule
\textbf{Model checkpoints} & \textbf{Layers} & \textbf{Hidden units} & \textbf{Heads} & \textbf{Total params} & \textbf{Batch size} & \textbf{Epochs}\\
\midrule
bert-base-uncased & 12&768 &12 & 110M & 32 &5 \\
deberta-base &12 & 768 & 12 & 100M & 32 &5 \\
distilbert-base& 6 & 768 & 12 & 66M & 32 & 5\\
\midrule
roberta-large-mnli & 24 & 1024 & 16 & 355M & 32 & 5\\
bert-base-uncased-MNLI & 12 & 768 & 12 & 110M & 32 & 10\\
deberta-base-mnli & 12 & 768 & 12 & 86M & 32 & 10\\
\bottomrule
\end{tabular}
}
\end{table}

For all transformer models, hyperparameter tuning is performed by exploring different combinations of learning rates, batch sizes, and training epochs. Batch sizes ranging from 8 to 128 and training epochs between 2 and 10 are evaluated using an adaptive learning-rate scheduler. Across all datasets, we observe that smaller batch sizes consistently provide better generalization than larger batches, while the optimal number of training epochs typically lies between five and ten. Early stopping based on the validation loss is employed to prevent overfitting and to retain the best-performing model checkpoint. All experiments are conducted on an NVIDIA P100 GPU equipped with 32 GB of memory.

The computational cost of fine-tuning largely depends on both the dataset size and the number of model parameters. Among the evaluated checkpoints, \textit{roberta-large-mnli} requires approximately 1--2 hours of training, whereas the lighter models, namely \textit{deberta-base-mnli} and \textit{distilbert-base}, typically complete training within 10--15 minutes. The \textit{bert-base-uncased-MNLI} checkpoint requires approximately 20--25 minutes to converge.

For the proposed SR-BERT framework, we initialize the sentence encoder using the pretrained \textit{distilbert-base-nli-mean-tokens}\footnote[7]{https://huggingface.co/sentence-transformers/distilbert-base-nli-mean-tokens} checkpoint, which is subsequently fine-tuned on the software requirement datasets to learn domain-specific sentence representations.

The datasets used in this study exhibit different class distributions. The CDN dataset contains three classes (Conflict, Duplicate, and Neutral) with relatively balanced class proportions, making it suitable for 5-fold cross-validation. In contrast, the remaining datasets consist of only Conflict and Neutral classes, with conflict pairs representing a small minority of the data. Consequently, 3-fold cross-validation is employed for these datasets. Table~\ref{tab:fold_sample_dist} presents the class distributions for the training and testing partitions in each cross-validation fold.

\begin{table}[!t]
\centering
\caption{Class distribution in each fold's training and test sets for all the datasets.}
\label{tab:fold_sample_dist}
\resizebox{\linewidth}{!}{
\begin{tabular}{lccccc|cccc}
\toprule
 & & \multicolumn{4}{c}{\textbf{Training}} & \multicolumn{4}{c}{\textbf{Test}} \\
\cmidrule(lr){3-6}\cmidrule(lr){7-10}
\textbf{Dataset} & \textbf{\# folds} & \textbf{\# D} & \textbf{\# C} & \textbf{\# N} & \textbf{Total} & \textbf{\# D} & \textbf{\# C} & \textbf{\# N} & \textbf{Total}\\
\midrule
CDN & 5 & 1,339 & 4,442 & 2,720 & 8,501 & 334 & 1,111 & 680 & 2,125 \\
\midrule
 & & \multicolumn{4}{c}{\textbf{Training}} & \multicolumn{4}{c}{\textbf{Test}} \\
\cmidrule(lr){3-6}\cmidrule(lr){7-10}
\textbf{Dataset} & \textbf{\# folds} & & \textbf{\# C} & \textbf{\# N} & \textbf{Total} & & \textbf{\# C} & \textbf{\# N} & \textbf{Total} \\
\midrule
CN & 3 & & 3,702 & 2,206 & 5,908 & & 1,851 & 1,134 & 2,985 \\
UAV & 3 & & 12 & 4,434 & 4,446 & & 6 & 2,218 & 2,224 \\
WorldVista & 3 & & 23 & 7,229 & 7,252 & & 12 & 3,614 & 3,626 \\
PURE & 3 & & 14 & 1,460 & 1,474 & & 6 & 731 & 737 \\
OPENCOSS & 3 & & 7 & 4,517 & 4,524 & & 3 & 2,259 & 2,262 \\
\bottomrule
\end{tabular}
}
\end{table}

For the proposed SR-BERT framework, each requirement pair is represented by three embedding vectors that are concatenated before being provided to the downstream classifier. Specifically, the representations correspond to the embeddings of the two individual requirements together with their element-wise difference. To provide a fair comparison, we also evaluate three baseline feature representations, namely TF-IDF, pooled GloVe--FastText embeddings, and SBERT embeddings. The corresponding feature extraction settings, SVM parameters, and SR-BERT hyperparameters are summarized in Table~\ref{tab:s_bert_parameters}.

\renewcommand{\arraystretch}{1.3}
\begin{table*}[!t]
\centering
\caption{Baseline embeddings with linear classifier (SVM) and SR-BERT hyperparameters.}
\label{tab:s_bert_parameters}
\resizebox{0.7\linewidth}{!}{
\begin{tabular}{|l|rrrrr|}
\hline
& {\bf max\_features} & {\bf ngram\_range} & {\bf Min\_df} & {\bf Vector length} & \\
\textbf{TF-IDF} & 1500 & (1,3) & 5 & 1380 & \\
& {\bf Embedding Size} & {\bf Context Window Size} & {\bf Max Sentence Length} & {\bf Embedding Type} & {\bf Vector length} \\
\textbf{FastText} & 4096 & 10 & 250 & LSTM-based & 500 \\
\textbf{GloVe} & 300 & - & - & Static & 300 \\
\hline
& {\bf C} & {\bf Gamma} & {\bf Kernel} & {\bf Degree} & {\bf Tolerance} \\
\textbf{SVM} & 1.0 & `auto' & Linear & 3 & 0.001 \\
\hline
& {\bf Loss function} & {\bf Epochs (max)} & {\bf Learning rate (LR)} & {\bf LR Scheduler} & {\bf Batch size}\\
\textbf{SBERT embeddings} & Softmax and cosine similarity loss & 3 & 1e-3 & Exponential & 16 \\
\textbf{SR-BERT embeddings} & Softmax loss & 5 & 1e-5 & Exponential & 16 \\
\hline
& {\bf Dense units} & {\bf Epochs (max)} & {\bf Activation function} & {\bf Dropout} & {\bf Batch size} \\
\textbf{Feed-forward neural network} & 1500 & 50 & ReLU & 0.2 & 16 \\
\hline
\end{tabular}
}
\end{table*}
\section{Results} \label{sec:results}
\subsection{CDN dataset classification results}
We compare SR-BERT, sequential fine-tuning checkpoints, and vanilla BERT against three baseline embeddings (TF-IDF, GloVe+FastText, and SBERT) on the CDN dataset using 5-fold cross-validation. Table~\ref{tab:biloss} reports weighted and macro precision, recall, and F1-scores.
SR-BERT obtains the highest weighted F1-score (95.3\%), followed by the sequentially fine-tuned MNLI checkpoints (\textit{deberta-base-mnli}: 92.6\%, \textit{bert-base-uncased-MNLI}: 92.0\%). The MNLI-adapted checkpoints improve over their vanilla counterparts, and the low standard deviations indicate stable performance across folds.
\renewcommand{\arraystretch}{1.2}
\begin{table*}[!t]
    \centering
    \caption{Results for CDN dataset. Values are averaged over 5 folds and results are shown as ``mean $\pm$ standard deviation''.}
    \resizebox{0.85\textwidth}{!}{{
    \label{tab:biloss} 
    \input{Tables/rq1_results.tex}
}}
\end{table*}
Table~\ref{tab:CDN_Class_specific} presents class-specific performance for the best model. SR-BERT achieves high recall for conflict and neutral pairs, while duplicate pairs have lower precision, recall, and F1-score. This pattern is consistent with the smaller duplicate class and the semantic overlap between duplicate and conflict pairs.
\begin{table}[!t]
\centering
\caption{Class-specific performance with the best performing model, SR-BERT, over CDN dataset. Reported values are averaged over 5 folds and results are shown as ``mean $\pm$ standard deviation''.}
\label{tab:CDN_Class_specific}
\resizebox{\linewidth}{!}{
    \input{Tables/best_cdn_res}
}
\end{table}

\subsection{Conflict and neutral classification results}
We evaluate binary conflict-neutral classification using 3-fold cross-validation due to smaller dataset sizes and severe class imbalance. Table~\ref{tab:CN_results} reports baseline, MNLI fine-tuned, and SR-BERT performance, including conflict-specific metrics.
\begin{table*}[!t]
\centering
    \caption{Performance values for conflict and neutral classification task for all the requirement pair datasets. Reported values are averaged over 3 folds and results are shown as ``mean $\pm$ standard deviation''. Dashed (-) values indicate 0\% as the performance value. The highest performance values are indicated by bold text and highlighted in blue for the Macro F1-score, and in yellow for the conflict class F1-score.}
    \label{tab:CN_results}
    \resizebox{0.85\linewidth}{!}{
    \input{Tables/cn_task_res}
}
\end{table*}
On CN, the MNLI-adapted checkpoints and SR-BERT all achieve high conflict-class F1-scores, with \textit{deberta-base-mnli} reaching the highest Conflict-F1 (99.6\%) and \textit{bert-base-uncased-MNLI} reaching the highest Macro-F1 (99.4\%). On the open-source datasets, the baseline embeddings and SR-BERT mostly predict the majority neutral class, leading to Macro-F1 values near 49\% and zero reported Conflict-F1. In contrast, the sequentially fine-tuned models retain conflict-detection ability under severe imbalance: \textit{bert-base-uncased-MNLI} reaches Macro-F1 scores of 90.8\% on WorldVista and 84.1\% on OPENCOSS, while \textit{deberta-base-mnli} reaches 87.7\% on UAV and 94.8\% on PURE. These results indicate that NLI initialization is especially valuable for imbalanced binary conflict detection.
\subsection{Cross-domain transfer learning results}
We evaluate cross-domain transfer for binary conflict classification using 3-fold cross-validation with macro-averaged metrics (Table~\ref{Tab:cross_domain_results}). Models trained on CN achieve high macro recall but low macro precision and very low Conflict-F1 on UAV, WorldVista, PURE, and OPENCOSS, indicating many false positives. This pattern suggests that CN's requirement structure does not transfer cleanly to the open-source datasets, so we exclude CN from the multi-source combinations.
\begin{table*}[!t]
\centering
    \caption{Evaluation of cross-domain models trained with different combinations of requirement pair datasets (Source Model). Reported values are averaged over 3 folds and presented as ``mean ± standard deviations". Improved results compared to Table~\ref{tab:CN_results} are underlined. The highest performance in terms of Macro F1-score and Conflict-F1 is highlighted in blue. The CN-trained model utilizes the transformer checkpoint \textit{deberta-base-mnli}, while the remaining configurations in the ``Source Model'' column are trained with \textit{bert-base-uncased-MNLI}}
    \label{Tab:cross_domain_results}
    \resizebox{0.7\linewidth}{!}{
   \input{Tables/cross_domain}
}
\end{table*}
We find that the strongest source configuration depends on the target dataset. PURE transfers best to UAV (Macro-F1: 89.2\%, Conflict-F1: 78.6\%); the combined UAV+PURE source performs best on WorldVista (Macro-F1: 93.6\%, Conflict-F1: 87.3\%), exceeding the in-domain WorldVista Macro-F1 of 90.8\%; and the combined UAV+WorldVista+OPENCOSS source performs best on PURE (Macro-F1: 89.8\%). For OPENCOSS, UAV alone achieves the best cross-domain Macro-F1 (67.7\%), suggesting that some source-target pairs share more transferable structure than broader multi-source combinations.

\subsection{Comparison with Existing Techniques}
We compare our approach against three baseline methods from the literature, ensuring fair evaluation by using identical datasets and folds where applicable.
\paragraph{Comparison with \citet{malik2025supervised}} Table~\ref{tab:baseline_compare} reports results on the WorldVista, UAV, PURE, and OPENCOSS datasets. The sequentially fine-tuned models improve over their supervised semantic similarity approach (S3CDA) on all four datasets. The largest gains occur on WorldVista (\textit{deberta-base-mnli}: 0.93 vs. 0.86) and OPENCOSS (\textit{bert-base-uncased-MNLI}: 0.74 vs. 0.43), suggesting that transformer-based transfer learning is useful when requirements contain general-domain terminology or when the dataset is severely imbalanced (OPENCOSS: 10 conflict vs. 6,776 neutral pairs). On PURE and UAV, where cosine similarity already separates the classes more clearly, the margins are narrower but still favor the sequentially fine-tuned models.
\begin{table}[!t]
    \centering
    \caption{Performance comparison with \citet{malik2025supervised} (supervised semantic similarity). Reported values are averaged over 3 folds as mean $\pm$ std.}
    \label{tab:baseline_compare}
    \resizebox{\linewidth}{!}{
    \begin{tabular}{l|c|c|c}
    \toprule
    \textbf{Dataset} & \textbf{deberta-base-mnli} & \textbf{bert-base-uncased-MNLI} & \textbf{\citet{malik2025supervised}} \\
    \midrule
    WorldVista & \textbf{0.93 $\pm$ 0.01} & 0.90 $\pm$ 0.03 & 0.86 $\pm$ 0.05\\
    UAV & 0.84 $\pm$ 0.12 & \textbf{0.94 $\pm$       0.01} & 0.90 $\pm$ 0.04 \\
    PURE & \textbf{0.90 $\pm$ 0.05} & 0.89 $\pm$ 0.05 & 0.88 $\pm$ 0.11 \\
    OPENCOSS & 0.69 $\pm$ 0.14 & \textbf{0.74 $\pm$ 0.17} & 0.43 $\pm$ 0.04 \\
    \bottomrule
    \end{tabular}
    }
\end{table}

\paragraph{Comparison with \citet{abeba2021identification}} Table~\ref{tab:baseline_compare_1} shows that their BiLSTM + Word2vec approach struggles with class imbalance, achieving near-random macro F1-scores (0.47--0.48) by predicting the majority class. In contrast, sequential fine-tuning of \textit{bert-base-uncased-MNLI} yields higher macro F1-scores (0.79--0.86), indicating that pretrained transformers with NLI-based initialization are more robust to imbalanced RE data than recurrent architectures with static embeddings in these experiments.
\begin{table}[!t]
    \centering
    \caption{Performance comparison (Macro F1) with \citet{abeba2021identification} (BiLSTM + Word2vec).}
    \label{tab:baseline_compare_1}
    \resizebox{\linewidth}{!}{
    \begin{tabular}{l|r|r|r}
    \toprule
    \textbf{Approach} & \textbf{WorldVista} & \textbf{PURE} & \textbf{UAV} \\
    \midrule
    \citet{abeba2021identification} -- BiLSTM + Word2vec & 0.47 & 0.48 & 0.48 \\
    bert-base-uncased-MNLI (Sequential) & \textbf{0.86} & \textbf{0.82} & \textbf{0.79} \\
    \bottomrule
    \end{tabular}
    }
\end{table}

\paragraph{Comparison with \citet{shah2021detecting}} Using a sample of 122 hardware requirements provided by the authors, our cross-domain model (trained on WorldVista/UAV/PURE) achieves 0.60 F1 versus 0.48 for their clustering-based approach (Table~\ref{tab:baseline_comp_2}). The 19-point recall gap (0.56 vs. 0.37) indicates that the cross-domain transformer recovers more conflict pairs in this sample, despite the domain difference between the training and test data.
\begin{table}[!t]
    \centering
    \caption{Performance comparison with \citet{shah2021detecting} (clustering-based) on 122 hardware requirements.}
    \label{tab:baseline_comp_2}
    \resizebox{\linewidth}{!}{
    \begin{tabular}{l|r|r|r}
    \toprule
    \textbf{Approach} & \textbf{Precision} & \textbf{Recall} & \textbf{F1-score} \\
    \midrule
    \citet{shah2021detecting} & 0.77 & 0.37 & 0.48 \\
    bert-base-uncased-MNLI (Cross-domain) & \textbf{0.83} & \textbf{0.56} & \textbf{0.60} \\
    \bottomrule
    \end{tabular}
    }
\end{table}

Taken together, these comparisons show that transfer learning with NLI-pretrained transformers performs better than the evaluated rule-based feature extraction, static-embedding recurrent, and clustering-based approaches on the tested datasets. The gains are most pronounced on small or imbalanced datasets (OPENCOSS and \citeauthor{shah2021detecting}'s sample) and on the dataset with more general-domain language (WorldVista), supporting the use of NLI-to-requirement transfer for this task.

\subsection{Discussion}\label{sec:discussion}
\begin{figure}[!t]
    \centering
    \includegraphics[width=0.5\textwidth]{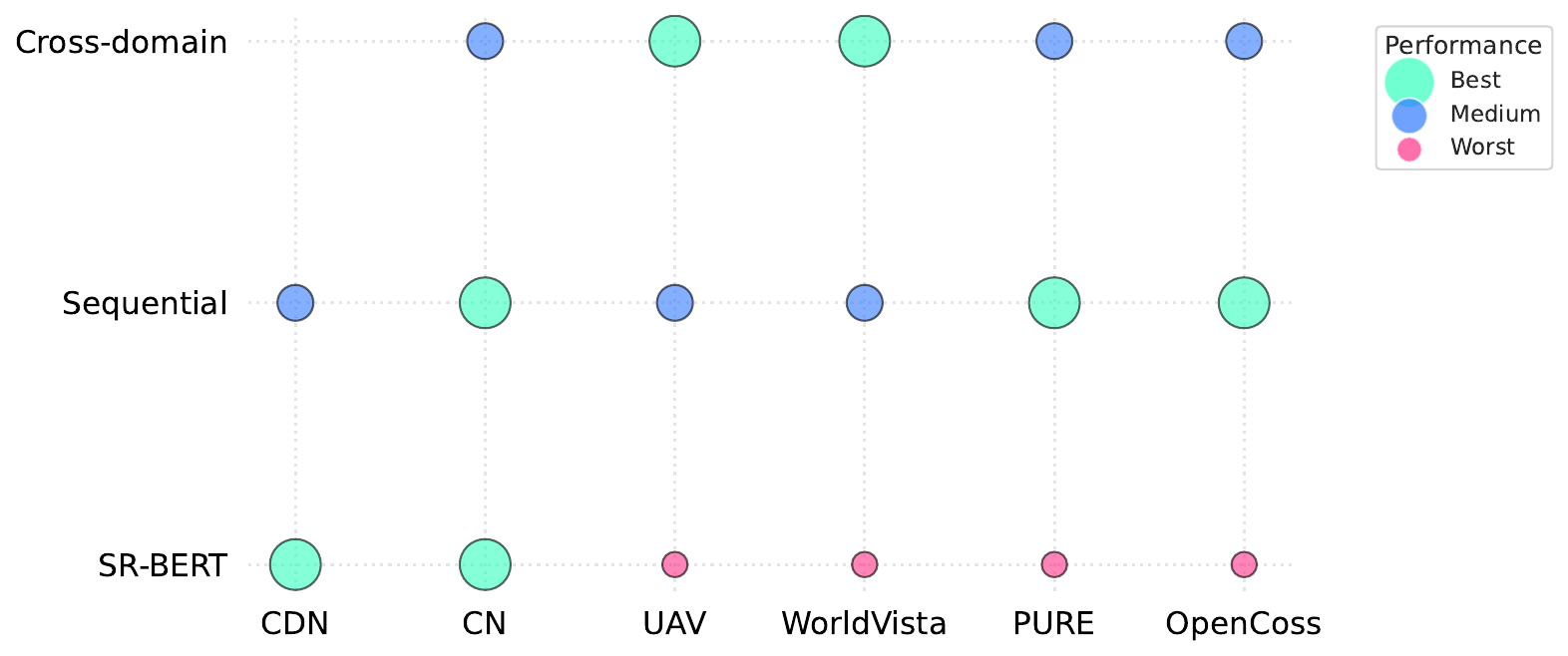}
    \caption{Summary comparison for requirement pair datasets. `Sequential' represents multi-stage FT. Bubble color indicates performance level, while size represents performance metric.}
    \label{fig:summary}
\end{figure}

Figure~\ref{fig:summary} compares top-performing techniques across datasets. SR-BERT and sequential methods perform best on CDN, where the larger training set supports learning conflict and duplicate patterns. CN's binary conflict-neutral setup also supports strong in-domain performance. In contrast, SR-BERT performs poorly on WorldVista, UAV, PURE, and OPENCOSS, where conflict pairs are sparse. Sequential transfer is strongest on PURE and OPENCOSS, whereas cross-domain transfer is strongest on UAV and WorldVista. This variation suggests that transfer performance depends not only on dataset size but also on requirement style, label balance, and source-target similarity.

Overall, the results suggest a dataset-dependent modeling choice: sequential transfer is preferable for small or imbalanced datasets, SR-BERT is effective when larger labeled requirement-pair data are available, and cross-domain transfer is useful when the source and target requirements share structural similarity. Generalization therefore depends on requirement documentation style, conflict prevalence, and domain proximity between datasets.
\section{Conclusion}\label{sec:conclusion}
This study investigated transformer-based methods for conflict and duplicate detection in software requirements. We formulated the problem as requirement-pair classification, evaluated sequential and cross-domain transfer learning with pretrained transformers, and proposed SR-BERT as a domain-adapted Sentence-BERT framework for this task.
The results show that SR-BERT achieves the strongest performance on the larger three-class CDN dataset, whereas sequential transfer learning with MNLI-initialized checkpoints is more effective for the smaller and more imbalanced binary conflict-neutral datasets. Cross-domain transfer is beneficial in selected source-target combinations, especially when requirement structure appears similar across datasets. Overall, the study supports sentence-pair transfer learning as a useful approach for automating requirements conflict and duplicate detection, while showing that performance remains sensitive to class balance and domain similarity.
The analysis is limited to six datasets, and evaluation on additional requirement corpora would be needed to strengthen generalizability claims. The scarcity of public conflict and duplicate datasets also required modifications to existing corpora, which should be considered a threat to validity. Future work includes data augmentation for rare conflict classes, evaluation of newer transformer checkpoints, stronger entity and role extraction for requirement semantics, and broader testing in domains with similarly high-stakes requirement documentation.







\small
\bibliographystyle{IEEEtranN} 
\bibliography{IEEEabrv,Ref}
\end{document}

%% file: Tables/rq1_results.tex
\begin{tabular}{l|cccccc}
\toprule
 & \multicolumn{3}{c}{\textbf{Weighted}} & \multicolumn{3}{c}{\textbf{Macro}} \\
\cmidrule(lr){2-4}\cmidrule(lr){5-7}
Models & Precision & Recall & F1 & Precision & Recall & F1 \\
\midrule
TF-IDF + SVM & 0.808 $\pm$ 0.013 & 0.807 $\pm$ 0.004 & 0.756 $\pm$ 0.006 & 0.804 $\pm$ 0.024 & 0.643 $\pm$ 0.005 & 0.629 $\pm$ 0.008 \\
WE~(Glove+FastText) + SVM  &0.591 $\pm$ 0.006  & 0.530 $\pm$ 0.003  &0.376 $\pm$ 0.007 &0.505 $\pm$ 0.006 & 0.341 $\pm$ 0.003 & 0.246 $\pm$ 0.007 \\
SBERT+linear layer &0.593 $\pm$ 0.018  & 0.622 $\pm$ 0.017 & 0.588 $\pm$ 0.016 &0.586 $\pm$ 0.018  & 0.594 $\pm$ 0.035 & 0.571 $\pm$ 0.024 \\
\midrule
bert-base-uncased &0.797 $\pm$ 0.049 & 0.833 $\pm$ 0.007 & 0.779 $\pm$ 0.019&0.749 $\pm$ 0.106 & 0.675 $\pm$ 0.026 & 0.646 $\pm$ 0.045\\
distilbert-base & 0.854 $\pm$ 0.006 & 0.866 $\pm$ 0.006 & 0.854 $\pm$ 0.007 & 0.813 $\pm$ 0.011 & 0.784 $\pm$ 0.012 & 0.794 $\pm$ 0.010 \\
deberta-base &0.812 $\pm$ 0.048  & 0.848 $\pm$ 0.013 & 0.812 $\pm$ 0.033 & 0.758 $\pm$ 0.089 & 0.715 $\pm$ 0.056 & 0.706 $\pm$ 0.067\\
\midrule
roberta-large-mnli & 0.923 $\pm$ 0.011 & 0.916 $\pm$ 0.011 & 0.918 $\pm$ 0.011 & 0.883 $\pm$ 0.013 & 0.907 $\pm$ 0.017 & 0.892 $\pm$ 0.014 \\
bert-base-uncased-MNLI & 0.923 $\pm$ 0.011 & 0.919 $\pm$ 0.008 & 0.920 $\pm$ 0.009 & 0.890 $\pm$ 0.007 & 0.901 $\pm$ 0.020 & 0.894 $\pm$ 0.011 \\
deberta-base-mnli & 0.928 $\pm$ 0.003 & 0.924 $\pm$ 0.002 & 0.926 $\pm$ 0.002 & 0.895 $\pm$ 0.002 & 0.909 $\pm$ 0.006 & 0.901 $\pm$ 0.002 \\
\midrule 
\rowcolor{cyan!15}\textbf{SR-BERT} & \textbf{0.953 $\pm$ 0.005} & \textbf{0.953 $\pm$ 0.004} & \textbf{0.953 $\pm$ 0.005} & \textbf{0.936 $\pm$ 0.003} & \textbf{0.938 $\pm$ 0.010} & \textbf{0.937 $\pm$ 0.006} \\ 
\bottomrule
\end{tabular}

%% file: Tables/best_cdn_res.tex
\begin{tabular}{l|cccr}
    \hline
     & \textbf{Precision} & \textbf{Recall} & \textbf{F1} & \textbf{support} \\ 
    \hline
    \textbf{Conflict}     & 0.956 $\pm$ 0.011 & 0.956 $\pm$ 0.003 & 0.956 $\pm$ 0.004 & 1110.60       \\ 
    \textbf{Duplicate}    & 0.855 $\pm$ 0.015 & 0.871 $\pm$ 0.032 & 0.862 $\pm$ 0.012 & 334.60        \\
    \textbf{Neutral}      & 0.997 $\pm$ 0.002 & 0.989 $\pm$ 0.005 & 0.993 $\pm$ 0.002 & 680.00        \\
    \hline
\end{tabular}

%% file: Tables/cn_task_res.tex
\begin{tabular}{llcccccc}
\toprule
 & & \multicolumn{3}{c}{\textbf{Macro}} & \multicolumn{3}{c}{\textbf{Conflict Class}} \\
\cmidrule(lr){3-5}\cmidrule(lr){6-8}
\textbf{Dataset} & \textbf{Models} & \textbf{Precision} & \textbf{Recall} & \textbf{F1} & \textbf{Precision} & \textbf{Recall} & \textbf{F1} \\
\midrule
\texttt{CN} & TF-IDF + SVM & 0.955 $\pm$ 0.001  & 0.931 $\pm$ 0.003 & 0.940 $\pm$ 0.002 & 0.926 $\pm$ 0.005 & 0.991 $\pm$ 0.003 & 0.957 $\pm$ 0.001 \\
 & WE(GloVe + FastText) + SVM  & 0.626 $\pm$ 0.223  & 0.522 $\pm$ 0.016 & 0.431 $\pm$ 0.034 &0.631 $\pm$ 0.007 &0.997 $\pm$ 0.001 &0.773 $\pm$ 0.005 \\
 & SBERT + Linear~layer &0.696 $\pm$ 0.049 & 0.595 $\pm$ 0.013& 0.579 $\pm$ 0.012 & 0.672 $\pm$ 0.006 & 0.935 $\pm$ 0.027& 0.782 $\pm$ 0.013 \\
 \cmidrule{2-8}
& \textbf{deberta-base-mnli} &  0.994 $\pm$ 0.000    & 0.992 $\pm$ 0.001 & 0.993 $\pm$ 0.001 & 0.993 $\pm$ 0.000 & 0.999 $\pm$ 0.000& \cellcolor{yellow!15}\textbf{0.996 $\pm$ 0.000} \\
& \textbf{bert-base-uncased-MNLI} &     0.995 $\pm$ 0.001    & 0.993 $\pm$ 0.001 & \cellcolor{cyan!15}\textbf{0.994 $\pm$ 0.001} & 0.991 $\pm$ 0.000 & 0.999 $\pm$ 0.000 & 0.995 $\pm$  0.000\\
& SR-BERT &0.994 $\pm$ 0.000  &0.991 $\pm$ 0.001  & 0.992 $\pm$ 0.001  &0.989 $\pm$ 0.001 & 0.999 $\pm$ 0.000 &0.994 $\pm$0.000 \\
\midrule
\texttt{UAV} & TF-IDF + SVM &0.498 $\pm$ 0.000  & 0.500 $\pm$ 0.000 & 0.499 $\pm$ 0.000 & - &- & -\\
& WE(GloVe + FastText) + SVM & 0.498 $\pm$ 0.000 & 0.500 $\pm$ 0.000 & 0.499 $\pm$ 0.000 & -&-&-\\
& SBERT + Linear~layer & 0.498 $\pm$ 0.000 & 0.499 $\pm$ 0.000 & 0.499 $\pm$ 0.000 &- & -&- \\
 \cmidrule{2-8}
 & \textbf{deberta-base-mnli} &    0.924 $\pm$ 0.058   & 0.860 $\pm$ 0.078 & \cellcolor{cyan!15}\textbf{0.877 $\pm$ 0.034} & 0.849 $\pm$ 0.117 & 0.722 $\pm$ 0.157 & \cellcolor{yellow!15}\textbf{0.756 $\pm$ 0.068} \\
& \textbf{bert-base-uncased-MNLI} &     0.876 $\pm$ 0.017    & 0.833 $\pm$ 0.067 & 0.849 $\pm$ 0.036 & 0.754 $\pm$ 0.035 & 0.666 $\pm$ 0.136 & 0.698 $\pm$ 0.071\\ 
& SR-BERT & 0.498 $\pm$ 0.000  & 0.500 $\pm$ 0.000 & 0.499 $\pm$ 0.000 & -&- &- \\
\midrule
\texttt{WorldVista} & TF-IDF + SVM &0.498 $\pm$ 0.000 & 0.500 $\pm$ 0.000 & 0.499 $\pm$ 0.000 &- &- &- \\
& WE(GloVe + FastText) + SVM  &0.498 $\pm$ 0.000 & 0.500 $\pm$ 0.000 & 0.499 $\pm$ 0.000 &- &- &- \\
& SBERT + Linear~layer &0.498 $\pm$ 0.000 & 0.498 $\pm$ 0.000 & 0.498 $\pm$ 0.000 &- &- &- \\
 \cmidrule{2-8}
  &\textbf{deberta-base-mnli}   & 0.999$\pm$ 0.000    & 0.827 $\pm$ 0.041 & 0.893 $\pm$ 0.031 & 1.000 $\pm$ 0.000  & 0.654 $\pm$ 0.083 & 0.787 $\pm$ 0.062  \\
& \textbf{bert-base-uncased-MNLI}          & 0.978 $\pm$ 0.029   &0.857 $\pm$ 0.051 &\cellcolor{cyan!15}\textbf{0.908 $\pm$ 0.043} & 0.958 $\pm$ 0.058 & 0.714 $\pm$ 0.102 & \cellcolor{yellow!15}\textbf{0.817 $\pm$ 0.087} \\ 
& SR-BERT &0.498 $\pm$ 0.000  & 0.500 $\pm$ 0.000 &0.499 $\pm$ 0.000 &- &- &- \\
\midrule
\texttt{PURE} & TF-IDF + SVM &0.495 $\pm$ 0.000 & 0.500 $\pm$ 0.000 &0.479 $\pm$ 0.000 & - & -& -  \\
& WE(GloVe + FastText) + SVM &0.495 $\pm$ 0.000   & 0.500 $\pm$ 0.000 & 0.497 $\pm$ 0.000 &- &- &- \\
& SBERT + Linear~layer &0.495 $\pm$ 0.000 &0.497 $\pm$ 0.000 & 0.496 $\pm$ 0.000 &-&- &- \\
 \cmidrule{2-8}
 & \textbf{deberta-base-mnli} &  0.971 $\pm$ 0.039    & 0.928 $\pm$ 0.058 & \cellcolor{cyan!15}\textbf{0.948 $\pm$ 0.048} & 0.944 $\pm$ 0.078 & 0.857 $\pm$ 0.116 & \cellcolor{yellow!15}\textbf{0.897 $\pm$ 0.095} \\
& \textbf{bert-base-uncased-MNLI} & 0.947 $\pm$ 0.036    & 0.903 $\pm$ 0.030 & 0.919 $\pm$ 0.005 & 0.896 $\pm$ 0.073 & 0.801 $\pm$ 0.062 & 0.841 $\pm$ 0.011\\ 
& SR-BERT &0.495 $\pm$ 0.000  & 0.500 $\pm$ 0.000 & 0.497 $\pm$ 0.000 & - &- &- \\
\midrule
\texttt{OPENCOSS} & TF-IDF + SVM &0.499 $\pm$ 0.000 & 0.499 $\pm$ 0.000 & 0.499 $\pm$ 0.000 &- & -&- \\
& WE(GloVe + FastText) + SVM & 0.499 $\pm$ 0.000 & 0.500 $\pm$ 0.000 & 0.499 $\pm$ 0.000 &- &- &- \\
& SBERT + Linear~layer &0.495 $\pm$ 0.000 &0.497 $\pm$ 0.000 & 0.496 $\pm$ 0.000 &-&- &-  \\
 \cmidrule{2-8}
 & \textbf{deberta-base-mnli} & 0.721 $\pm$ 0.157  & 0.777 $\pm$ 0.207 & 0.744 $\pm$ 0.175 & 0.444 $\pm$ 0.314  & 0.555 $\pm$ 0.415 & 0.488 $\pm$ 0.349 \\
&\textbf{bert-base-uncased-MNLI} & 0.958 $\pm$ 0.058   & 0.791 $\pm$ 0.089  & \cellcolor{cyan!15}\textbf{0.841 $\pm$ 0.065} & 0.916 $\pm$ 0.117 & 0.583 $\pm$ 0.180  & \cellcolor{yellow!15}\textbf{0.683 $\pm$ 0.131}\\ 
& SR-BERT &0.499 $\pm$ 0.000 & 0.500 $\pm$ 0.000 &0.499 $\pm$ 0.000 &- &- &- \\
\bottomrule
\end{tabular}

%% file: Tables/cross_domain.tex
\begin{tabular}{ll|ccc|c}
\toprule
 \textbf{Target Data} &  \textbf{Source Model} &  \textbf{Precision} &  \textbf{Recall} &  \textbf{F1} & \textbf{Conflict-F1} \\ 
\midrule
\texttt{UAV} &  CN & 0.509 $\pm$ 0.000 & 0.926 $\pm$ 0.002    & 0.478 $\pm$ 0.002  & 0.035 $\pm$ 0.001          \\
& WorldVista & 0.592 $\pm$ 0.002 & 0.966 $\pm$ 0.038& 0.651 $\pm$ 0.001 &0.309 $\pm$ 0.002 \\
& PURE &0.872 $\pm$ 0.040 &0.916 $\pm$0.068 & \underline{\cellcolor{cyan!15}\textbf{0.892 $\pm$ 0.052}} & \underline{\cellcolor{cyan!15}\textbf{0.786 $\pm$ 0.105}} \\
& OPENCOSS &0.498 $\pm$ 0.000 &0.500 $\pm$ 0.000 &0.499 $\pm$ 0.000 & -\\
& WorldVista + PURE &0.876 $\pm$ 0.119 & 0.777 $\pm$ 0.103 & 0.798 $\pm$ 0.083 & 0.598  $\pm$ 0.166 \\
  &  WorldVista + PURE + OPENCOSS  & 0.991 $\pm$ 0.000 &  0.666 $\pm$ 0.068   & 0.741 $\pm$ 0.078  &  0.484  $\pm$ 0.155        \\ 
\midrule
\texttt{WorldVista} & CN & 0.527 $\pm$ 0.002 & 0.959 $\pm$ 0.018   & 0.538 $\pm$ 0.005 & 0.105 $\pm$ 0.009          \\
& UAV &0.999 $\pm$ 0.000 &0.702 $\pm$ 0.049 &0.783 $\pm$ 0.048 & 0.568 $\pm$ 0.097 \\
& PURE & 0.978 $\pm$ 0.029 & 0.843 $\pm$ 0.036 & \underline{\textbf{0.899 $\pm$ 0.035}} & \underline{\textbf{0.799 $\pm$ 0.070}} \\
& OPENCOSS &0.498 $\pm$ 0.000 & 0.500 $\pm$ 0.000 & 0.499 $\pm$0.000 &- \\
& UAV + PURE &0.946 $\pm$ 0.047 & 0.929 $\pm$ 0.017& \underline{\cellcolor{cyan!15}\textbf{0.936 $\pm$ 0.031}} & \underline{\cellcolor{cyan!15}\textbf{0.873  $\pm$ 0.062}} \\
  & UAV + PURE + OPENCOSS   & 0.498 $\pm$ 0.000       & 0.500 $\pm$ 0.000     & 0.499 $\pm$ 0.000&-         \\
\midrule
\texttt{PURE} &  CN & 0.522 $\pm$ 0.002 & 0.904 $\pm$ 0.008 & 0.491 $\pm$ 0.010 &  0.087 $\pm$ 0.010 \\
& UAV &0.970 $\pm$ 0.038 & 0.829 $\pm$ 0.084 & 0.876 $\pm$ 0.055 &0.755 $\pm$ 0.109  \\
& WorldVista &0.783 $\pm$ 0.048 & 0.996 $\pm$ 0.001 & 0.857 $\pm$ 0.042 & 0.718 $\pm$ 0.083\\
& OPENCOSS &0.495 $\pm$ 0.000 & 0.500 $\pm$ 0.000 & 0.497 $\pm$ 0.000 &- \\
& UAV + WorldVista &0.998 $\pm$ 0.000 & 0.829 $\pm$ 0.061 & 0.893 $\pm$ 0.043 & 0.787 $\pm$ 0.085\\
 & UAV + WorldVista + OPENCOSS         & 0.965 $\pm$ 0.047    & 0.848 $\pm$ 0.011   & {\cellcolor{cyan!15}0.898 $\pm$ 0.025} &   {\cellcolor{cyan!15} 0.797 $\pm$ 0.049}\\ 
\midrule
\texttt{OPENCOSS}    &  CN           & 0.501 $\pm$ 0.000 & 0.642 $\pm$ 0.001    & 0.223 $\pm$ 0.002  &   0.004 $\pm$ 0.000      \\
 & UAV &0.647 $\pm$ 0.026 & 0.749 $\pm$ 0.067 & {\cellcolor{cyan!15} 0.677 $\pm$ 0.015} & {\cellcolor{cyan!15} 0.355 $\pm$ 0.031} \\
 & WorldVista &0.508 $\pm$ 0.000 & 0.955 $\pm$ 0.002 &0.492 $\pm$ 0.000 & 0.031 $\pm$ 0.002\\
 & PURE &0.522 $\pm$ 0.002 &0.984 $\pm$0.001 & 0.535 $\pm$ 0.004 & 0.087 $\pm$ 0.008 \\
 &WorldVista + PURE &0.565 $\pm$ 0.006  & 0.940  $\pm$ 0.077 & 0.611 $\pm$ 0.011 & 0.226 $\pm$ 0.023\\
 &  UAV + WorldVista + PURE  & 0.515 $\pm$ 0.001       & 0.976 $\pm$ 0.002   &  0.517 $\pm$ 0.002 & 0.059 $\pm$ 0.005   \\ 
 \bottomrule
\end{tabular}